\documentclass{ws-procs9x6}

\newcommand{\ii}{\mathrm{i}}

\newcommand{\singlefig}{.7\textwidth}
\newcommand{\doublefig}{0.95\textwidth}

\begin{document}

\title{Charge transport in a
 nonlinear, three--dimensional DNA model  with disorder}

\author{JFR Archilla}
\address{Nonlinear Physics Group of the University of Sevilla,
Departamento de F\'{i}sica Aplicada I, ETSI Inform\'atica, Avda
Reina Mercedes s/n, 41012-- Sevilla, Spain\\Email: archilla@us.es}
\author{D Hennig and J Agarwal}
\address{Freie Universit\"{a}t Berlin, Fachbereich Physik,
Institut f\"{u}r Theoretische Physik, Arnimallee 14,
14195--Berlin, Germany}

\maketitle

\abstracts{We study the transport of charge due to polarons in a
model of DNA which takes in account its 3D structure and the
coupling of the electron wave function with the H--bond
distortions and the twist motions of the base pairs. Perturbations
of the ground states lead to moving polarons which travel long
distances. The influence of parametric and structural disorder,
due to the impact of the ambient, is considered, showing that the
moving polarons survive to a certain degree of disorder.
Comparison of the linear and tail analysis and the numerical
results makes possible to obtain further information on the moving
polaron properties.}

\section{Introduction}
 \label{sec:intro}
Charge transport along DNA is a subject of particular interest for
two main  reasons: on the one hand, it plays a fundamental role in
biological functions as repair and biosynthesis; on the other
hand, because of possible applications in molecular electronics
and as molecular wires~\cite{Ratner}. Results on experimental DNA
conductivity are controversial. It has been reported that it is
good conductor~\cite{Tran}, insulator~\cite{Braun} and
semiconductor~\cite{Porath}. A possible explanation, among others,
for these striking differences can be the different long-range
correlations of the DNA sequences~\cite{Carpena}.

In this work we propose a nonlinear mechanism for charge transport
along DNA in the framework of the base pair
picture~\cite{Cocco,Barbi}, taking into account its spatial
structure and the coupling of the spatial and electron variables.
It turns out to be an efficient mechanism for charge transport
which survives to a certain degree of diagonal and structural
disorder.

\section{Model}
\label{sec:model}

\begin{figure}[ht]
  \begin{center}
   \includegraphics[width=\singlefig]{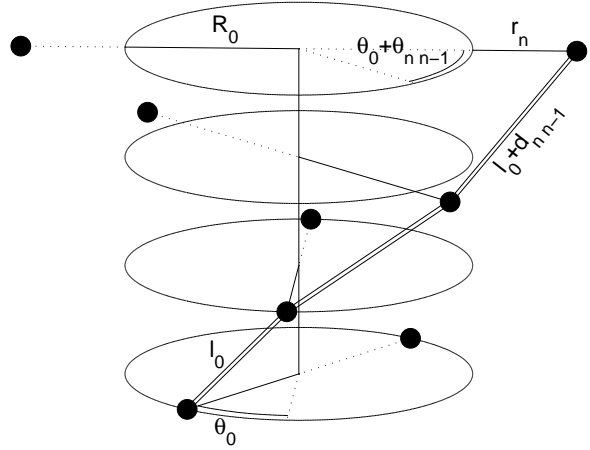}
  \caption{Sketch of the helical structure of
the DNA model, the bases being represented by bullets. Geometrical
parameters $R_0$, $\theta_0$, $l_0$ and the radial and angular
variables $r_n$ and $\theta_{n,n-1}$, respectively are indicated.
}
  \label{fig:sketch}
 \end{center}
\end{figure}

We consider a semi--classical three--dimensional, tight--binding
model for the DNA molecule, the sketch of the system being shown
in Fig.~\ref{fig:sketch}.
    The lattice oscillators are treated classically while the charge
is described by a quantum system. The justification is that the
nucleotides are large molecules with mass numbers of about $300$.
This also implies that  molecular motions are small and slow
compared to the one of the charge particle. The Hamiltonian of the
whole system is given by
 $\hat{H}=\hat{H}_{\mathrm{el}}+\hat{H}_{\mathrm{rad}}+
\hat{H}_{\mathrm{twist}}\, $,
 where $\hat{H}_{\mathrm{rad}}$ and
$\hat{H}_{\mathrm{twist}}$ are, {\em de facto}, classical
Hamiltonians and we can omit the hat on them. The Hamiltonians
corresponding to the distances between nucleotides in each base
pair $H_{\mathrm{r}}$, and the twist motion $H_{\mathrm{twist}}$
are given by
\begin{equation}
\label{eq:hrad}
 H_{\mathrm{rad}}=\sum_{n}
\frac{\left(p^{\mathrm{r}}_{n}\right)^2}{2\,M}
+\frac{M\,\Omega_{r}^2r_n^2}{2};\,\,\,
 H_{\mathrm{twist}}=\sum_{n}
\frac{\left(\,p^{\mathrm{\theta}}_{n,n-1}\right)^2}{2\,J}
+\frac{J\,\Omega_{\mathrm{\theta}}^2\theta_{n,n-1}^2}{2}\,
\end{equation}
where $\{r_n\}$ represent the stretchings from the equilibrium
distance, $M$ the reduced mass or each base pair,  $\Omega_{r}$
the linear radial frequency,  $\{\theta_{n,n-1}\}$ are the angles
between two consecutive base pairs with respect to their
equilibrium value $\theta_0=36^{\circ}$, $J$ is the inertia
moment, and $\Omega_\theta$, the twist linear frequency.

The electronic part is given by a tight--binding system of the
form
  \begin{eqnarray}
  \label{eq:helec}
  \hat{H}_{\mathrm{el}}=\sum_n E_n|n \rangle \langle n| -
  V_{n-1,n}|n-1\rangle \langle n|
-V_{n+1,n}|n+1\rangle \langle n|\,.
  \end{eqnarray}
In the state $|n \rangle $ the charge carrier is localized at the
$n^{\mathrm{th}}$ base pair. The quantities $V_{n-1,n}$ represent
the nearest--neighbour transfer integrals along the base pairs and
$E_n$ are the on--site matrix--elements. We write the electronic
state as $|\Psi\rangle = \sum_n c_n(t)|n\rangle $, where $c_n(t)$
is the probability amplitude of finding the electron at the state
$|n\rangle$. The interaction between the electronic variables and
the structure variables $r_n$ and $\theta_{n,n-1}$ arises from the
dependence of the electronic parameters $E_n$ and $V_{n,n-1}$ on
the spatial coordinates, given by  $E_n=E_n^0+k\,r_n$ and
$V_{n,n-1}=V_0\, (1-\alpha \,d_{n,n-1})$, where
 $d_{n,n-1}$ is the first order Taylor expansion of the distances
 between consecutive nucleotides, with respect to their equilibrium
 value. Is is given by
\begin{equation}
d_{n,n-1}\simeq \frac{R_0}{l_0}\left[ (1-\cos(
\theta_0))\,(r_n+r_{n-1})+\sin(\theta_0)\,
R_0\,\theta_{n,n-1}\right]\,. \label{eq:distance}
\end{equation}
 Realistic parameters for DNA
molecules are given in Refs.~\cite{Peyrard,Stryer,HAA02}.
 We scale the time according to
$t\rightarrow \Omega_{r}\,t$, which allow us to write the
Hamiltonian in terms of dymensionless quantities~\cite{HAA02}.
The classical Hamiltonian is defined as
$H_{\mathrm{class}}=\langle \Psi|\hat{H}|\Psi\rangle$, the
dynamical equations
 for $r_n$ and
$\theta_{n,n-1}$ are $\dot{p}_n^{\mathrm{r}}=
M\,\ddot{r}_n=-\partial H_{\mathrm{class}}/\partial r_n$ and
$\dot{p}_{n,n-1}^{\mathrm{\theta}}=
J\,\ddot{\theta}_{n,n-1}=-\partial H_{\mathrm{class}}/\partial
\theta_{n,n-1}$,  while the evolution equations for the electron
variables $c_n$ are obtained from the Schr\"odinger equation $\ii
\hbar\, (\partial \Psi/\partial
t)=\hat{H}_{\mathrm{el}}|\Psi\rangle$, which is equivalent to
$\dot{c}_n=-(\ii/\hbar)(\partial H_{\mathrm{class}}/\partial
c_n^*)$. In this way we can obtain the scaled dynamical equations,
which are
\begin{eqnarray}
\ii\,\tau\dot{c}_{n}&=&(E_n^0\,+k\,r_n)\,c_n\nonumber\\
&-&(1-\alpha\,d_{n+1,n})\,c_{n+1} -(1-\alpha\,d_{n,n-1})\,c_{n-1}
\label{eq:dotc}\\ \ddot{r}_{n}&=&-r_n-k\,|c_n|^2\,-\,\alpha\,
\frac{R_0}{l_0}\,(1-\cos \theta_0)\,\nonumber\\ &\times&
\left\{\,[c_{n+1}^*c_{n}+c_{n+1}c_{n}^*]+[c_{n}^*c_{n-1}+c_{n}c_{n-1}^*]\,
\right\}\label{eq:dotr}\\
\ddot{\theta}_{n,n-1}&=&-\Omega^2\,\theta_{n,n-1}\,-\,
\alpha\,V\,\frac{R_0^2}{l_0}\, \sin
\theta_0\,[c_{n}^*\,c_{n-1}\,+\,c_{n}\,c_{n-1}^*]\,,
\label{eq:dottheta}
\end{eqnarray}
where $\tau=\hbar\,\Omega_{r}/V_0$ represents the time scale
separation between the fast electron motion and the slow bond
vibrations.

The values of the scaled parameters are~\cite{HAA02}
$\tau=0.2589$, $\Omega^2=[0.709-1.417]\times 10^{-2}$, $V=0.0823$,
$R_0=34.862$ and $l_0=24.590$, the time unit being $\sim 1.6\,
\mathrm{ps}$. Note that the time scales for the different
variables
 differ in an order of magnitude: the fastest variables are the
$\{c_n\}$, with a characteristic frequency of order $1/\tau \sim
4$, followed by $\{r_n\}$ with the unity, and $\{\theta_{n,n-1}\}$
with $\Omega \sim 0.08$

There are no reliable values for the electron--radial and
electron--twist coupling parameters, $k$ and $\alpha$. The
criterion we have taken for their values is the consistency of the
 numerical simulations  with the hypothesis, i.e.,
the deformations of the helix are small, and the linear
approximations of the distance $d_{n,n-1}$ and the trigonometric
functions remain valid. Typical values are $k\sim 1$ and $\alpha
\sim 0.002 $.

\section{Stationary states}

We suppose initially that $r_n$ and $\theta_{n,n-1}$ are constant,
i.e., the Born-Oppenheimer approximation. This allows the
obtention of expressions for them from
Eqs.~(\ref{eq:dotr})--(\ref{eq:dottheta}) which are inserted in
Eq.~(\ref{eq:dotc}), leading to a nonlinear Schr\"odinger equation
for the electronic amplitudes:
\begin{eqnarray}
\ii\,\tau\,\dot{c}_{n}&=&\left[E_n^0-k^2\,|c_n|^2-k\,\alpha\,
\frac{R_0}{l_0}\,(1-\cos \theta_0)\right. \nonumber\\
&\times&\left.\left\{\,[c_{n+1}^*c_{n}+c_{n+1}c_n^*]
+[c_n^*c_{n-1}+c_n c_{n-1}^*]\,\right\}\,\right]\,c_n \nonumber\\
&-&(\,1-\alpha\,d_{n+1\,n}\,)\,c_{n+1}\, -\,(\,1-\alpha \,
d_{n,n-1}\,)\,c_{n-1} \label{eq:DNLS}\,,
\end{eqnarray}
whith $\{d_{n,n-1}\}$ given by Eq.~(\ref{eq:distance}), depending
algebraically on the $\{c_n\}$ through $\{r_n\}$ and
$\{\theta_{n,n-1}\}$.

To obtain stationary localized solutions we use a numerical
method~\cite{Kalosakas,Nick}. We substitute
$c_{n}=\Phi_{n}\,\exp(-\ii\,E\,t/\tau)$ in Eq.~(\ref{eq:DNLS}),
with $\Phi_n$ constants, and obtain a nonlinear difference system
$E\,\Phi=\hat{A}\,\phi$, with $\Phi=(\Phi_1,\dots,\Phi_N)$. Its
solutions are attractors of the map:
\begin{equation}
\Phi\rightarrow \Phi'= \hat{A}\,\Phi/\|\hat{A}\,\Phi\|\, ,
\label{eq:map}
\end{equation}
$\|\cdot\|$ being the quadratic norm. We start with a completely
localized state at a site $n_0$, i.e., $\Phi_n=\delta_{n,n_0}$ and
apply the map above until $\Phi'=\Phi$. In this way we obtain
stationary localized solutions and their energy $E$.

In the ordered case, with $E_n^0=E_0,\forall n$, $E_0$ can be made
zero with the ansatz $c_n\rightarrow c_n\exp(-\ii E_0 t/\tau)$. We
obtain symmetric polarons with  width of about $20$ sites,
combined with a local compression of order $\sim 0.15$ and a local
unwinding of $\sim 1^\circ$ and energies $E\lesssim -2\sim
-0.2\,\mathrm{eV}$. Considering static diagonal disorder with
random $E_n^0$ the localization is enhanced, due to Anderson
localization, the polaron being asymmetric, with a specific  shape
that depends on each particular disorder implementation. Similar
results are obtained with structural disorder, i.e., random
equilibrium values $R_n^0$ and $\theta_n^0$. The Floquet analysis
shows that these polarons are linearly stable~\cite{HAA02}.

\section{Charge transport}

To activate the polaron motion we need to perturb the zero
velocities of the ground state with a localized, spatially
antisymmetric mode. This is obtained either with the
time--consuming pinning--mode method~\cite{Chen}, either with the
simpler discrete gradient method, i.e., perturbing the variables
$r_n$ with velocities parallel to
$\{r_{n+1}-r_{n-1}\}$~\cite{IST02}. The main difference is that
the first guarantees the mobility, while the second does not, but
in practice works well very often. Values of the modulus of the
{\em kick} velocity $\lambda_\mathrm{r}=(\sum
\dot{r}_n^2(0))^{1/2} \sim 0.02$, equivalent to a kinetic energy
of $\sim 200\,\mathrm{meV}$ are appropriate to obtain good
mobility with low radiation.

\begin{figure}[ht]
  \begin{center}
   \includegraphics[width=\doublefig]{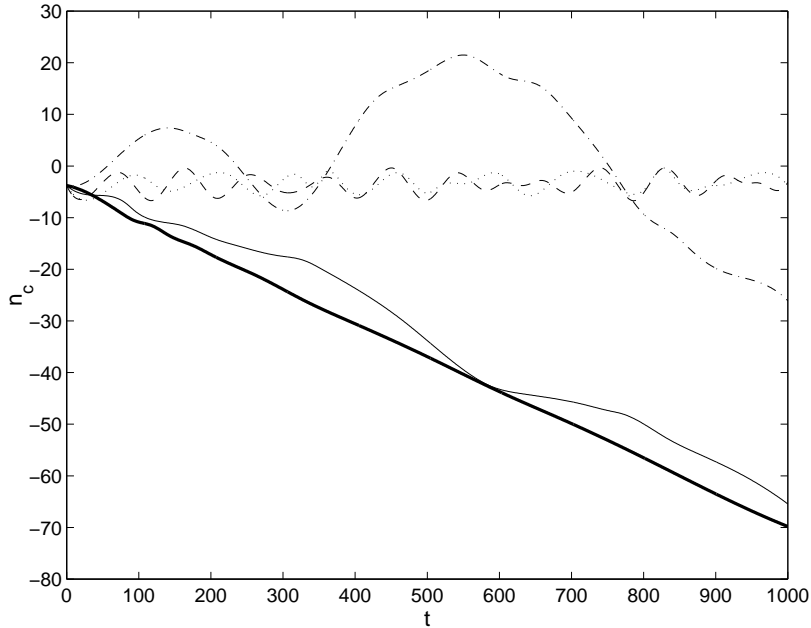}
  \caption{Diagonal
disorder. The position of the center of the electron breather as a
function of time and for different amounts of disorder. Full line
(Ordered case), short dashed line ($\Delta E=0.025$),
dashed-dotted line ($\Delta E=0.050$), dotted line ($\Delta
E=0.100 $) and long dashed line ($\Delta E=0.500$).}
\label{fig:disorderedmotion}
 \end{center}
\end{figure}

We obtain  propagation of the localized electron amplitudes with
constant velocity. A complementary compression and unwinding of
the helix travels with it, while a localized oscillation of the
angular variables remains at the initial positions. If we increase
the diagonal disorder with $E_n\in [-\Delta E, \Delta E]$, the
charge transport still takes place, until values of $\Delta E\sim
0.05$. Thereafter mobility becomes impossible and the polaron is
pinned to the lattice. With disorder, however, the velocity of the
first momentum of the electronic occupation amplitude
$n_c(t)=\sum_n n\,|c_n(t)|^2$ is not uniform, as shown in
Fig.~\ref{fig:disorderedmotion}.

We also consider structural disorder with random distributions of
the base pair spacings and angles around their ordered case
values, with different mean standard deviations $\Delta$. The
conclusion is that, up to values of $\Delta=10\%$, it does not
significantly affect polaron mobility, although there is a slight
reduction of its velocity.

It is interesting to remark that for the values of $\alpha$
coherent with our small amplitude hypotheses it is not possible to
move the polaron by {\em kicking} the angular variables. These
results are coherent with the fact that the Floquet analysis shows
that the pinning mode appears only in the radial variables and its
eigenvalue separates from the optimal value around $1$ when the
disorder is increased~\cite{HAA02}.
\section{Linear and tail analysis}
\label{sec:linearfloquet} Although our system is studied in the
nonlinear regime, and the results obtained throughout this work
are essentially nonlinear, there is a number of linear techniques
that can be applied as a reference and as a tool for obtaining
useful information. They are the study of the linear system itself
and the tail analysis.
\subsection{The linear system}
\label{sub:linear} If in Eqs.~(\ref{eq:dotc})--(\ref{eq:dottheta})
we cancel out the nonlinear terms, we obtain:
\begin{eqnarray}
\ii \,\tau\,\dot{c}_{n}=-c_{n+1}-c_{n-1} 
\,,\quad
 \ddot{r}_{n}=-r_n
 \,,\quad
\ddot{\theta}_{n\,n-1}=-\Omega^2\,\theta_{n\,n-1}\,.
\label{eq:dotlineal}
\end{eqnarray}
In the first equation the terms $E_n$, which are all equal to some
value $E_0$ in an homogeneous system, have been eliminated through
the ansatz $c_n \rightarrow c_n \exp (-\ii\,{E_0\,t}/{\tau})$. The
system becomes decoupled and the variables $r_n$ and
$\theta_{n\,n-1}$ correspond to independent linear oscillators
with frequencies $w_r=1$ and $\Omega=0.0842$, respectively, in the
scaled units. To obtain the ground state we substitute in
Eq.~(\ref{eq:dotlineal}) $c_n(t)=\phi_n \exp (-\ii\,E\,t/\tau)$,
$\phi_n$ being time independent. We obtain the stationary discrete
Schr\"odinger equation: $ E\,\phi_n=-\phi_{n+1}-\phi_{n-1}$.
Substitution of the linear modes $\phi_n=\exp(\ii\,q\,n)$ leads
to: $E=-2\,\cos q\, $. Therefore, the linear energy spectrum runs
from $-2$ to $2$. The minimum energy, $E=-2$, corresponds to the
wave vector $q=0$, i.e., all the oscillators vibrating in phase.
The nonlinear ground states described above have $E\lesssim -2$
and derive from this mode.
\subsection{Tail analysis}
\label{sub:tail} For sites far enough from the polaron center,
which can be only a few sites, we can still apply the linear
analysis. Now, the substitution of the tail mode $c_n(t)=\phi_n
\exp (-\xi \,n -\ii \,q\,n)$, with $n>0$ and $\xi>0$,  leads to:
 $E=-2\,\cos(q)\,\cosh (\xi)$ and $2\,\sin(q)\,\sinh(\xi)=0$.
The second equation, implies that only two wave vectors are
possible, $q=0$, with negative energy, and $q=\pi$, with positive
one. The energy of the first, which is the one that appears
throughout this paper because it is movable, is given by:
 $ E=-2\cos(\xi) < -2$.
 Therefore,
the nonlinearity, produces localized states with larger values of
the frequency $w_p=|E|/\tau$ outside the linear spectrum. The
distance of the energy calculated in the full system from
$E_{q=0}=-2$, is a measure of the degree of nonlinearity. It is
interesting to relate the values of the energies with the polaron
breadth. Here, we define it loosely
 as three times the number of
sites necessary for $|\phi_n|^2$ to decrease to a $5\%$. The
factor $3$ is there to allow for both and for the nonlinear center
of the polaron. This leads to $\Delta n\sim4.5/\xi$. An
approximate table is:
\begin{center}
\begin{tabular}{|l|c|c|c|c|c|} 
\hline
  $\xi$  & 2    & 1  &  0.5    & 0.15   & 0.075        \\ \hline
 $ \Delta n $ & 2.25  & 4.5  &  9      & 30     & 60            \\ \hline
  E      & -7   & -3 & -2.25   & -2.02  & -2.006        \\
 \hline
\end{tabular}
\end{center}
where the dimensionless $E$ units are equivalent to
$0.1\,\mathrm{eV}$. This makeshift method gives a good estimate of
the breadth of the excitation and produces a good fitting a few
units far from its center. The numerically calculated polarons
found in this work correspond to $\Delta n \sim 20$.

The tail analysis can also be applied to the moving polaron, far
enough from its center. Let us propose a localized traveling wave
of the form:
 $c_n=\exp\{ -\xi\,(n+v\,t) -\ii\,(q\,n + w_p \,t)\}$,
which represents a traveling localized wave moving to the left
behind the polaron, with positive $\xi$, $q$, and velocity $v$.
Substitution in Eq.~(\ref{eq:dotlineal}) leads to the following
equations: $E\equiv w_p \tau=-2\,\cos(q)\,\cosh(\xi)$ and
$\tau\,\xi\,v=2\,\sin(q)\,\sinh(\xi)$.
 They  are undetermined, as
there are four unknowns: $E$, $q$, $\xi$ and $v$, and do not allow
us to determine the velocity. This is little surprise, as the
polaron velocity depends on the energy given with the perturbation
to the static one, and this is done to the spatial variables.
However, these equations can be used to obtain an estimate of the
moving polaron energy and  wave number, using the decay length
$\xi$ corresponding to the static one, as there is no appreciable
change of shape, and the velocity observed within the simulations.
The conclusion is that there is only a negligible increase of the
energy, i.e., the energy given by the {\em kick} is stored in the
spatial variables. The fact that there is a imaginary part of the
energy $-i\,\xi \,v \,\tau$ is, as it is known, consequence of the
fact that we are dealing with only a part of the
system~\cite{Cohen}, in which the energy is decreasing as the
polaron moves to the left.

\section{Conclusions}
We have considered a model for charge transport along DNA, taking
into account the 3D structure. We obtain linearly stable,
localized stationary states. We are able to move them using the
pinning mode and the discrete gradient methods. For the ordered
case, the translational velocity is constant, while for the
disordered case the electron motion is irregular and if the
disorder is high enough it is impossible. We have performed a
linear and tail analysis of the system, which makes possible to
obtain the phonon spectrum, measure the degree of nonlinearity,
obtain estimates of the polaron breath and energy. It shows that
the charge transports very little energy, being most of it
associate with the spatial coordinates. The main conclusion of the
whole work can be that our proposed mechanism is an efficient
means for charge transport along ordered and disordered DNA.

\section*{Acknowledgments}
 The authors are grateful to partial support under the LOCNET EU network
HPRN-CT-1999-00163. JFRA  acknowledges DH and the
 Institut f\"{u}r Theoretische Physik for their warm hospitality

\enlargethispage{\baselineskip}
\newcommand{\noopsort}[1]{} \newcommand{\printfirst}[2]{#1}
  \newcommand{\singleletter}[1]{#1} \newcommand{\switchargs}[2]{#2#1}

\end{document}